\documentclass{iopart}
\usepackage{graphicx}
\usepackage{iopams}
\usepackage{psfrag}
\usepackage[]{caption2}
\usepackage{subfigure}
\newcommand{\ep}{\varepsilon}
\newcommand{\bsym}{\boldsymbol}
\begin{document}
\title{Soft modes in the easy plane pyrochlore antiferromagnet}
\author{J D M Champion\dag\ and P C W Holdsworth\ddag}

\address{\dag\ ISIS Facility, Rutherford Appleton Laboratory, Chilton,
Didcot, Oxon, OX11~0QX, UK.}
\address{\ddag\ Laboratoire de Physique, Ecole Normale Sup\'erieure, 46
All\'ee d'Italie, F-69364 Lyon, France.}

\ead{d.champion@rl.ac.uk}

\begin{abstract}

Thermal fluctuations lift the high ground state degeneracy of the
classical nearest neighbour pyrochlore antiferromagnet, with easy
plane anisotropy, giving a first order phase transition to a long
range ordered state. We show, from spin wave analysis and
numerical simulation, that even below this transition a continuous
manifold of states, of dimension $N^{2/3}$ exist ($N$ is the
number of degrees of freedom). As the temperature goes to zero a
further `order by disorder' selection is made from this manifold.
The pyrochlore antiferromagnet Er$_2$Ti$_2$O$_7$, is believed to
have an easy plane anisotropy and is reported to have the same
magnetic structure. The latter is perhaps surprising, given that
the dipole interaction lifts the degeneracy of the classical model
in favour of a different structure. We interpret our results in
the light of these facts.


\end{abstract}

In this paper we present a Monte Carlo study and classical spin
wave analysis of an easy plane antiferromagnet on a pyrochlore
lattice. The work was motivated by a recent comprehensive study of
the rare earth pyrochlore Erbium titanate~\cite{erbiumRC}.
Er$_2$Ti$_2$O$_7$, which orders at 1.2 K~\cite{Blote} has been
suggested to approximate the easy plane
antiferromagnet~\cite{Rosenkranz,Siddharthan,MMM}. We find that
the magnetic structure of this simple model system is determined
by an order by disorder mechanism~\cite{Villain} and that it is
the same structure as that observed in the experiment.

\section*{The easy plane antiferromagnet}

\begin{figure}
\begin{minipage}[t]{0.5\textwidth}
\centering
\includegraphics[width=0.99\textwidth]{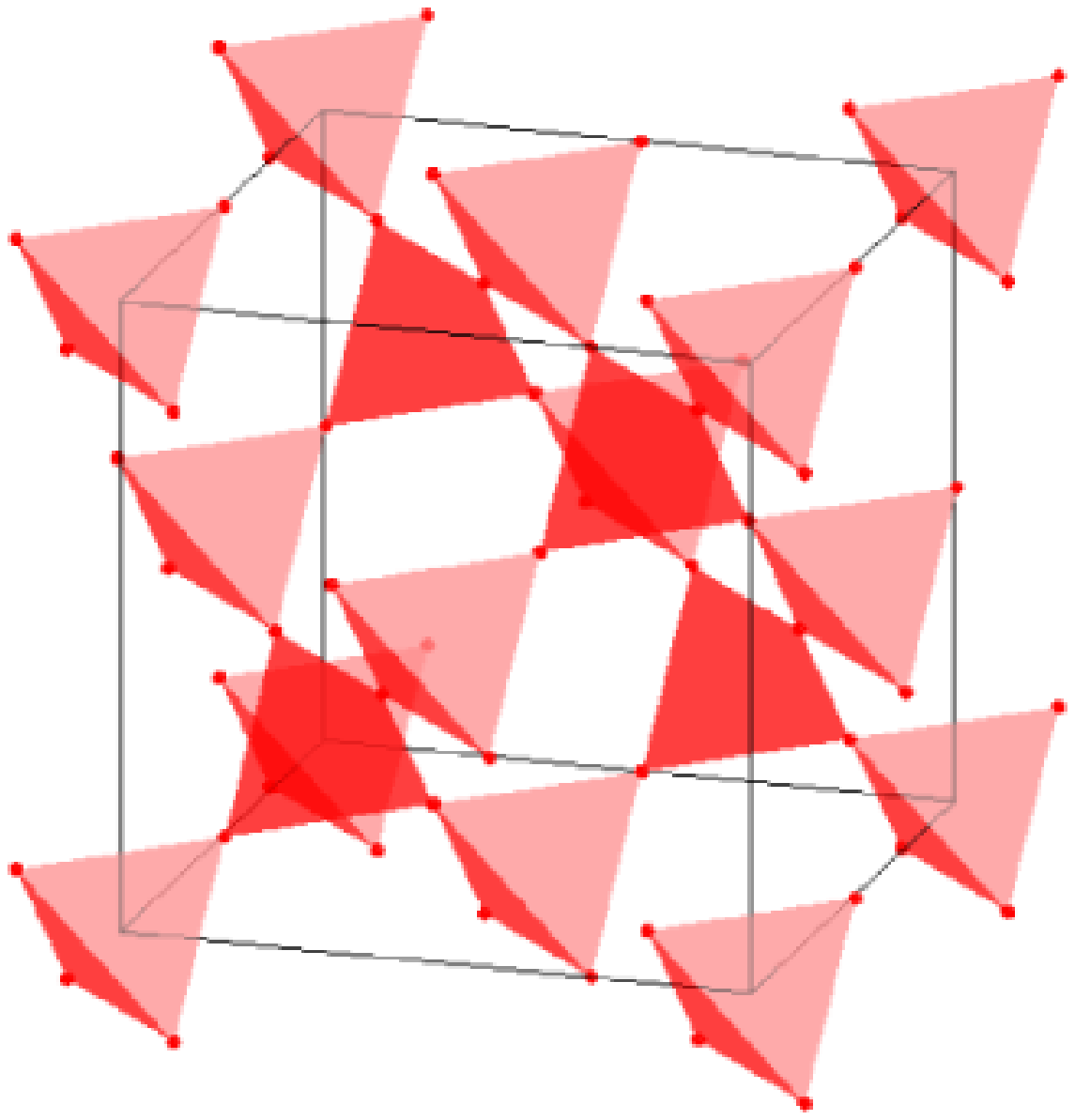} \caption{The pyrochlore lattice of corner-sharing tetrahedra}\label{pyro}
\end{minipage}
\hspace{0.01\textwidth}
\begin{minipage}[t]{0.5\textwidth}
\centering
\includegraphics[width=0.99\textwidth]{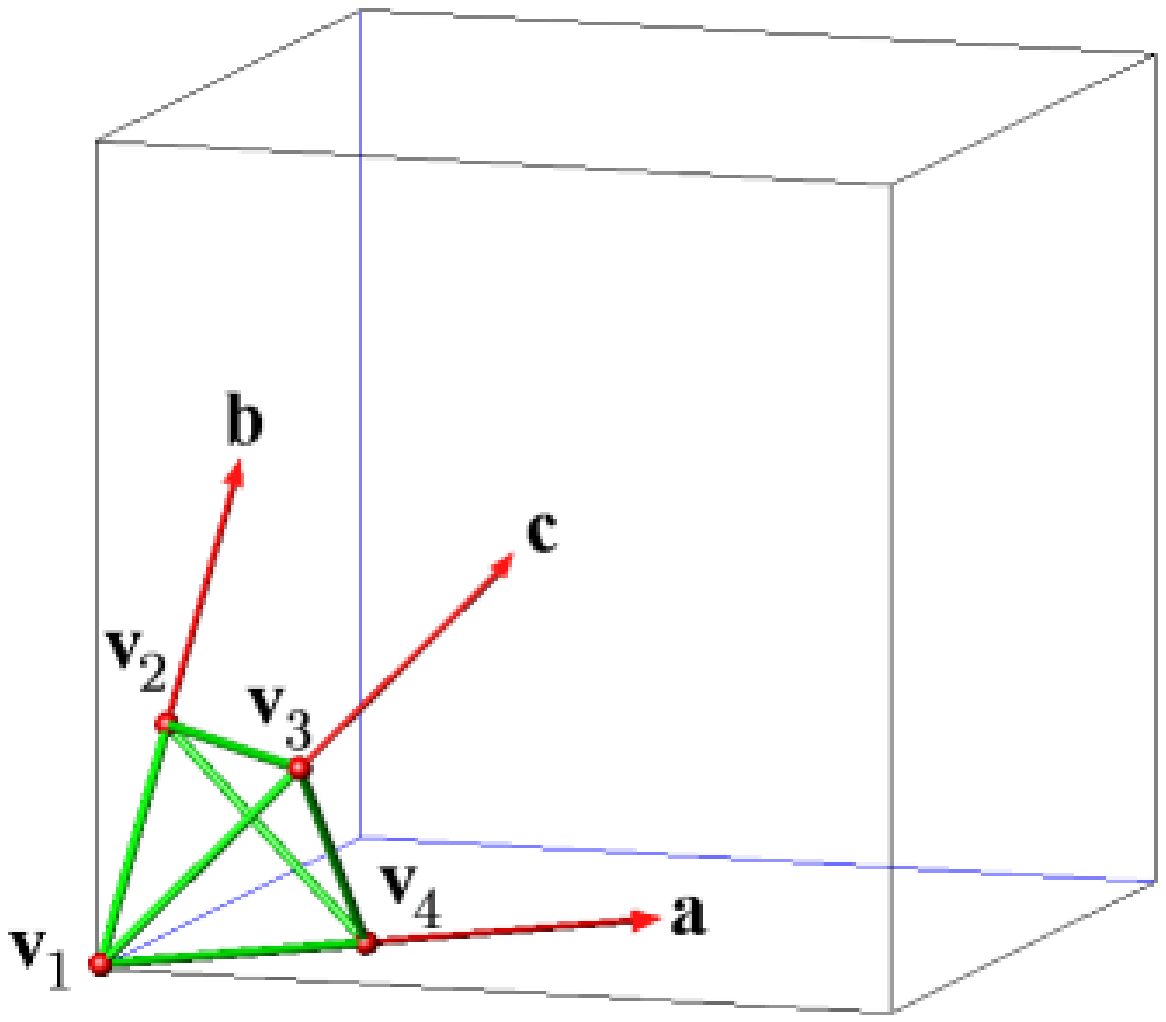} \caption{Rhombohedral axes with respect to the cubic unit
cell} \label{rhomucell}
\end{minipage}
\end{figure}

The pyrochlore lattice (see figure~\ref{pyro}), has a rhombohedral
primitive unit cell with lattice vectors,
\begin{eqnarray*}
\mathbf{a}=\left(\frac{1}{2}~\frac{1}{2}~0\right)\;,
\mathbf{b}=\left(\frac{1}{2}~0~\frac{1}{2}\right)\;,
\mathbf{c}=\left(0~\frac{1}{2}~\frac{1}{2}\right),
\end{eqnarray*}
 with a four atom basis at positions,
 $\mathbf{v}_{1}=0~0~0\; ,
\mathbf{v}_{2}=\frac{\mathbf{b}}{2}\; ,
\mathbf{v}_{3}=\frac{\mathbf{c}}{2}\; ,
\mathbf{v}_{4}=\frac{\mathbf{a}}{2}$, as shown in
figure~\ref{rhomucell}.
The units are those of a larger, cubic cell containing four
tetrahedral cells~\cite{RBS}. The number of primitive cells,
$N_{cell}=N/4$, where $N$ is the total number of spins and the
linear dimension of the sample , $L = (N_{cell}/4)^{1/3}$.

We have studied the $D\rightarrow - \infty$ limit of the
Hamiltonian
\begin{displaymath}
\mathbb{H} = -J\sum_{<i,j>} \mathbf{S}_i\cdot \mathbf{S}_j
-D\sum_{i=1}^{N}\left(\bsym{\delta}_i\cdot \mathbf{S}_i\right)^2,
\end{displaymath}
where $\bsym{\delta}_i$ corresponds to a local $<111>$ anisotropy
and the exchange constant $J=-1$. The spins are thus constrained
to a plane oriented perpendicular to the local $<111>$ axis:
 \numparts\begin{eqnarray}
\label{lin1}
S_{x1}+S_{y1}+S_{z1}=0\\
 S_{x2}-S_{y2}+S_{z2}=0 \\
 S_{x3}-S_{y3}-S_{z3}=0\\
 S_{x4}+S_{y4}-S_{z4}=0.
\end{eqnarray}\endnumparts
The ground state condition for four antiferromagnetically coupled
spins gives 3 further constraints~\cite{RBS}: \numparts
\begin{eqnarray}
S_{x1}+S_{x2}+S_{x3}+S_{x4}=0\\
S_{y1}+S_{y2}+S_{y3}+S_{y4}=0\\
\label{lin7}S_{z1}+S_{z2}+S_{z3}+S_{z4}=0.
\end{eqnarray}
\endnumparts
In addition the spins are all of unit length:
\begin{equation}
\label{nonlin}|\mathbf{S}_1|=|\mathbf{S}_2|=|\mathbf{S}_3|=|\mathbf{S}_4|=1.
\end{equation}

Linear equations (\ref{lin1})~--~(\ref{lin7}) can be solved by
row-echelon matrix reduction~\cite{BOAS}. The resulting dependent
and independent variables are put into equation (\ref{nonlin}) to
give four equations in terms of five variables. There is therefore
one independent and continuous degree of freedom for four spins on
a single tetrahedron in their ground state configuration. Full
details of this solution can be found in Ref.~\cite{PHD}.

The result for one tetrahedron suggests the possibility of a
continuous ground state degeneracy when the tetrahedra are
connected into a pyrochlore lattice and this is indeed what we
found numerically, as shown below. This result is rather different
from that found in Ref.~\cite{BGR}, where the model was first
studied. Here the authors only identify a discrete degeneracy and
it was a surprise to find evidence of fluctuations over a
continuous set of states.

\begin{figure}
\centering
\includegraphics[width=0.5\textwidth]{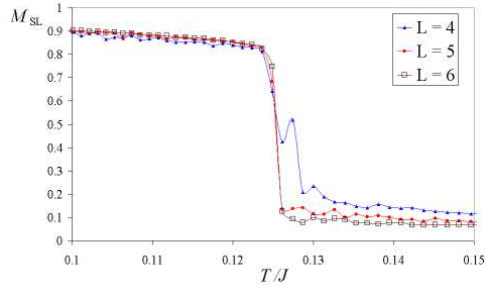}
\caption{$M_{SL}$ vs $T/J$, showing a first order phase
transition to an ordered q=0 state.}\label{slplot}
\end{figure}

\section*{Monte Carlo simulations}
Single spin-flip Monte Carlo simulations were performed on systems
of size $N=432$ to $N=3456$, with periodic boundaries. The
simulation lengths were typically $10^5$ moves per spin, with
$40,000$ moves used for equilibration. As shown in figure
\ref{slplot}, initial results confirm the strongly first order
phase transition presented in Ref.~\cite{BGR}. The order parameter
$M_{SL}$ is the sublattice magnetisation for each of the sites of
the primitive cell, which we refer to as $q=0$ ordering. For
$T\sim 0.125J$ the order parameter jumps from a small value
characteristic of a finite system in a disordered phase to
$M_{SL}\sim 0.85$, indicating an already highly ordered state.

\begin{figure}
\subfigure[State 1] {\label{GS1}
\begin{minipage}[b]{0.45\textwidth}
\centering \includegraphics[scale=0.3]{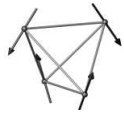}
\end{minipage}}
\hspace{1.0cm} \subfigure[State 2] {\label{GS2}
\begin{minipage}[b]{0.45\textwidth}
\centering \includegraphics[scale=0.37]{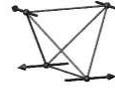}
\end{minipage}}
\caption{Different ground states of the easy-plane tetrahedron antiferromagnet.}
\end{figure}

Closer inspection~\cite{erbiumRC} shows that the ground state
chosen by the system is that shown in figure \ref{GS1}, not that
proposed in Ref.~\cite{BGR} and shown in figure \ref{GS2}. The,
initially proposed state has pairs of spins either aligned
anti-parallel to each other or perpendicular to each other. We
actually find scalar products between spins here of $-2/3$ and
$1/3$. The spin components of both states are shown in
Table~\ref{components}.

\Table{\label{components}The spin components of State 1 and State
2.}\br &\centre{3}{State 1}&\centre{3}{State 2}\\
\ns i&$S_x$&$S_y$&$S_z$&$S_x$&$S_y$&$S_z$\\
\mr
1&$-\frac{1}{\sqrt{6}}$&$-\frac{1}{\sqrt{6}}$&$\frac{2}{\sqrt{6}}$&
$\frac{-1}{\sqrt{2}}$&$\frac{1}{\sqrt{2}}$&0
\\
2&$\frac{1}{\sqrt{6}}$&$-\frac{1}{\sqrt{6}}$&$-\frac{2}{\sqrt{6}}$&
$-\frac{1}{\sqrt{2}}$&$-\frac{1}{\sqrt{2}}$&0
\\
3&$-\frac{1}{\sqrt{6}}$&$\frac{1}{\sqrt{6}}$&$-\frac{2}{\sqrt{6}}$&
$\frac{1}{\sqrt{2}}$&$\frac{1}{\sqrt{2}}$&0
\\
4&$\frac{1}{\sqrt{6}}$&$\frac{1}{\sqrt{6}}$&$\frac{2}{\sqrt{6}}$&
$\frac{1}{\sqrt{2}}$&$-\frac{1}{\sqrt{2}}$&0
\\
\br
\end{tabular}
\end{indented}
\end{table}

It is easy to show that not all possible ground states have $q=0$
ordering. For example, in Ref.~\cite{BGR} O($L^2$) states were
identified with spins ordered along one axis and with disorder in
the planes perpendicular to this axis. The transition must
therefore be driven by an order by disorder~\cite{Villain}
mechanism, although no details of the entropic forces involved
have previously been given. Below the transition it is clear that
further entropic selection between different $q=0$ states must
occur as we have already identified two different possibilities in
figures \ref{GS1} and \ref{GS2}, while the system systematically
ends up in state 1 for $T\rightarrow 0$.

\begin{figure}
\subfigure[$T/J=0.09$] {\label{j0-09}
\begin{minipage}[b]{0.3\textwidth}
\psfrag{P}{\huge{$P(E_{ij})$}}
 \psfrag{E}{\huge{$E_{ij}$}}
 \psfrag{0}{\huge$-1$}
 \psfrag{800}{\huge$~~1$}
 \psfrag{400}{\huge$~~0$}
\psfrag{1}{\huge$0$} \setlength{\abovecaptionskip}{1pt}
\includegraphics[angle=-90,width=3.9cm]{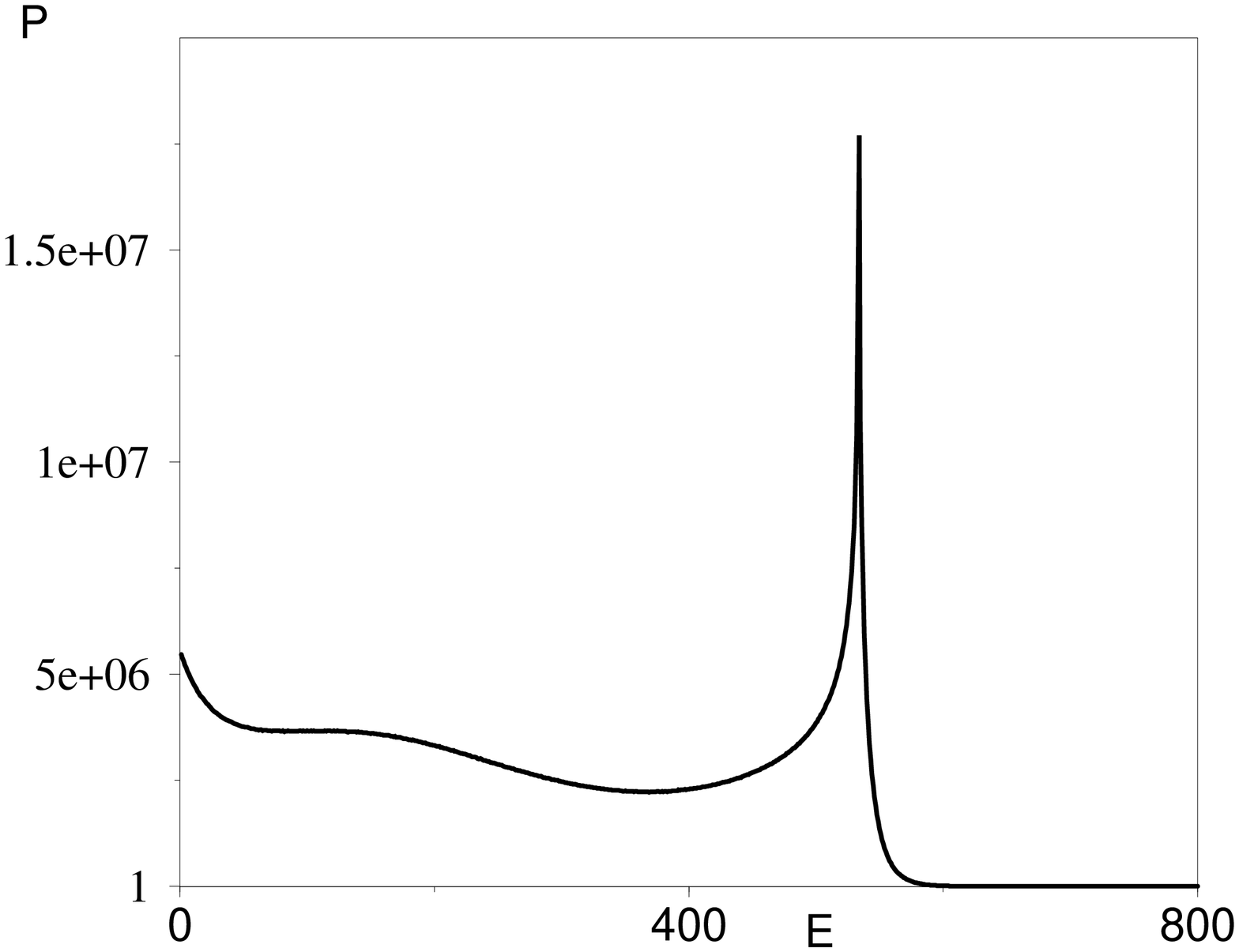}
\end{minipage}}
\hspace{0.2cm} \subfigure[$T/J=0.05$] {\label{j0-05}
\begin{minipage}[b]{0.3\textwidth}
\psfrag{P}{\huge{$P(E_{ij})$}}
 \psfrag{E}{\huge{$E_{ij}$}}
 \psfrag{0}{\huge$-1$}
 \psfrag{800}{\huge$~~1$}
 \psfrag{400}{\huge$~~0$}
\psfrag{1}{\huge$0$} \setlength{\abovecaptionskip}{1pt}
\includegraphics[angle=-90,width=3.9cm]{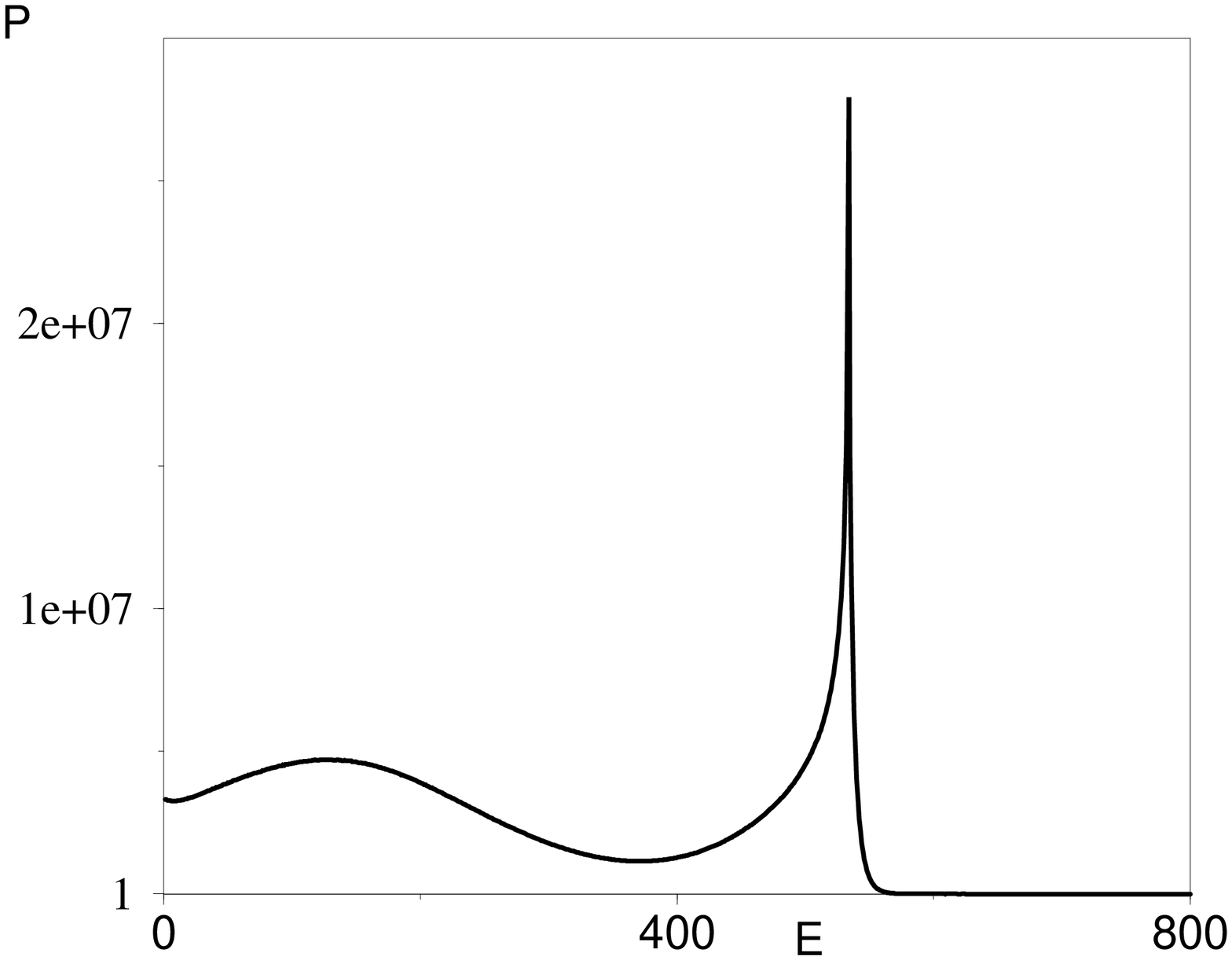}
\end{minipage}}
\hspace{0.2cm} \subfigure[$T/J=0.005$] {\label{j0-005}
\begin{minipage}[b]{0.3\textwidth}
\psfrag{P}{\huge{$P(E_{ij})$}}
 \psfrag{E}{\huge{$E_{ij}$}}
 \psfrag{0}{\huge$-1$}
 \psfrag{800}{\huge$~~1$}
 \psfrag{400}{\huge$~~0$}
\psfrag{1}{\huge$0$} \setlength{\abovecaptionskip}{1pt}
\includegraphics[angle=-90,width=3.9cm]{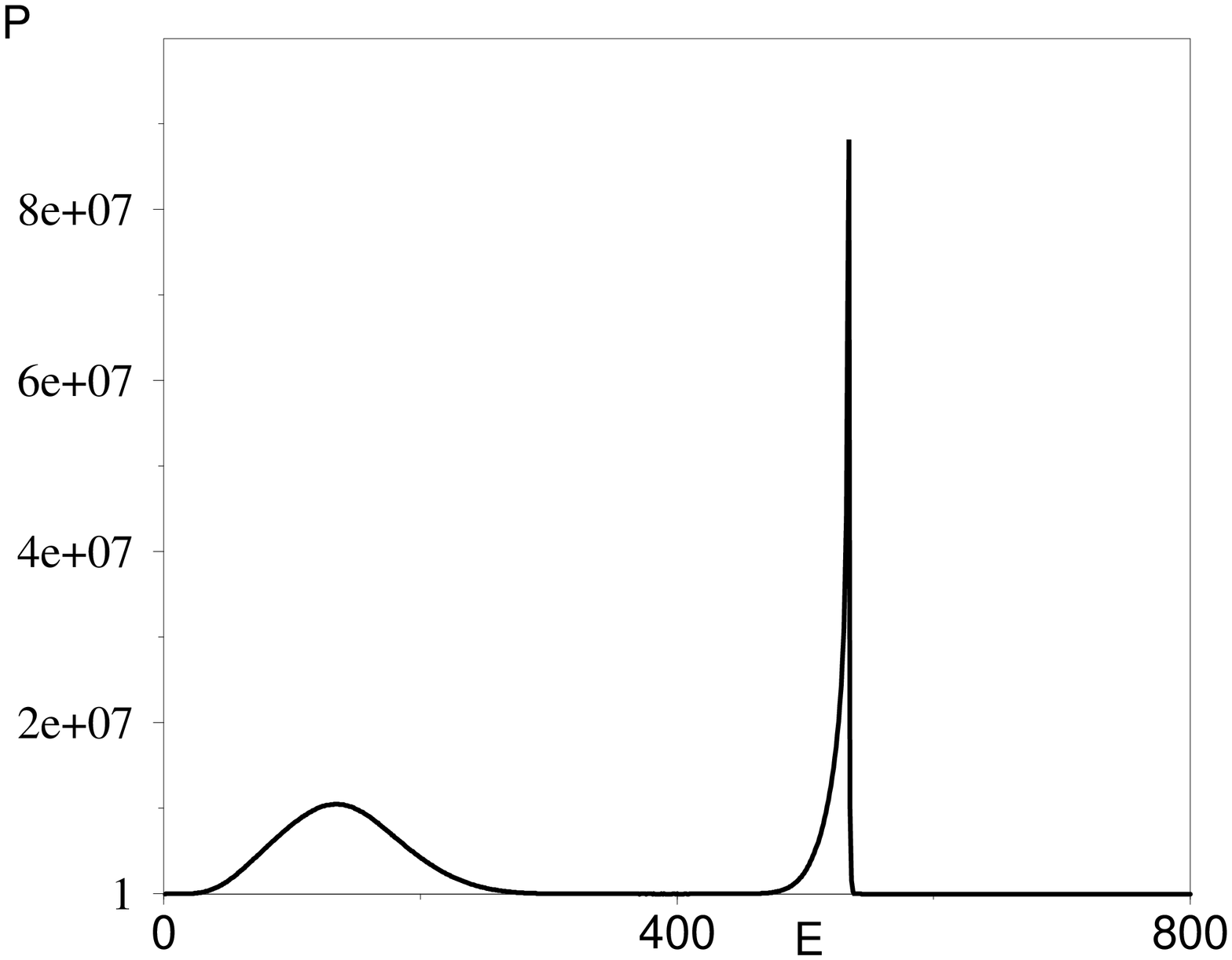}
\end{minipage}}
\caption{Bond energy distributions $P(E_{ij})$ for various $T/J$.}
\end{figure}

In figures \ref{j0-09}, \ref{j0-05} and \ref{j0-005} we illustrate
the evolution of this choice between different $q=0$ states. We
show the distribution of nearest neighbour bond energies for three
temperatures below the transition, for which, in all cases $M_{SL}
>0.9$. For the highest temperature there is a continuous
distribution of values from $-1$ (antiparallel spins) to $1/3$
where a sharp peak occurs. This figure is consistent with the
result for a single tetrahedron, that a continuous degeneracy of
$q=0$ ground states exists. One can see that going to the lowest
temperatures the distribution slowly becomes peaked around
energies $-2/3$ and $1/3$, corresponding to the ground state shown
in figure~\ref{GS1}. At all temperatures the energy is very close
to the ground state energy and this figure clearly represents the
order by disorder selection of a single point on a continuous
energy surface. We expect the selection to be due to a singular
excitation spectrum above the ground state at this
point~\cite{CHS} and this is what we show in the next section.

\section*{Spin wave calculations}

We consider small fluctuations out of the ground state 1. Working
in the laboratory reference frame and re-defining
$^{\alpha}\mathbf{S}, \; \alpha=1,4$ as the spins in a given
primitive cell, we make a small rotation,
${^{\alpha}S_x}={^{\alpha}S_{x0}}+{^{\alpha}\ep}$ compatible with
the local planar constraint. $S_{x0}$ is the ground state
component and $^{\alpha}\ep$ is the independent variable.
$^{\alpha} S_y(^{\alpha}\ep)$ and $^{\alpha} S_z(^{\alpha}\ep)$
can thus be calculated by application of the
constraints~\cite{PHD}. The Hamiltonian is then expanded to order
$(^{\alpha}\ep)^2$ and written in the form
\begin{equation}\label{dynham}
\mathbb{H}-H_o=\frac{1}{2}\sum_i^{N_{cell}}\sum_j^{N_{cell}}
\sum_{\alpha=1,\beta=1}^{\alpha=4,\beta=4}
{^{\alpha}\ep_i}\,\,\,M^{\alpha\beta}_{ij}\,\,\,{{^\beta}\ep_j},
\end{equation}
where $\alpha$ and $~\beta$ denote spins in a unit cell and $i$
and $j$ denote different unit cells (since all bonds are counted
twice the sum is divided by 2). Note that due to the frustration,
individual bonds make linear contributions in $(^{\alpha}\ep)$,
but as the ground state is a global minimum, these sum to zero.
Once the dynamical matrix $M^{\alpha\beta}(\mathbf{R}_{ij})$ has
been determined, it is Fourier transformed such that :
\begin{equation}
M^{\alpha\beta}(\mathbf{q})=\frac{1}{N_{cell}}\sum_{\mathbf{R}_{ij}}
M^{\alpha\beta}(\mathbf{R}_{ij})\exp{(i\mathbf{q}.\mathbf{R}_{ij})},
\end{equation}
and~(\ref{dynham}) is reduced to
\begin{equation}
\mathbb{H}-H_o=\frac{1}{2}\frac{1}{N_{cell}}\sum_{\mathbf{q}}\sum_{\alpha\beta}
{^{\alpha}\ep_{\mathbf{q}}}\,\,M^{\alpha\beta}(\mathbf{q})\,\,{^{\beta}\ep_{\mathbf{q}}},
\end{equation}
where
\begin{equation}
{^{\alpha}\ep_{\mathbf{q}}}=\frac{1}{\sqrt{N_{cell}}}\sum_{\mathbf{R}_{i}}{^{\alpha}\ep_i}\exp{(-i\mathbf{q}.\mathbf{R}_{i})}.
\end{equation}

$M^{\alpha\beta}(\mathbf{q})$ is shown in
matrix~(\ref{matrixjdm}):
\begin{equation}\label{matrixjdm}
\fl\left (
\begin{array}{cccc}
 4J&0&0& 2J\left(1+\e^{\left(i\mathbf{q}.\mathbf{a}\right)}\right) \\
 0&4J&2J\left(1+\e^{\left(i\mathbf{q}.\left(\mathbf{c}-\mathbf{b}\right)\right)}\right)&0\\
 0&2J\left(1+\e^{\left(-i\mathbf{q}.\left(\mathbf{c}-\mathbf{b}\right)\right)}\right)&4J&0  \\
 2J\left(1+\e^{\left(-i\mathbf{q}.\mathbf{a}\right)}\right)&0&0&4J\\
\end{array}
\right).
\end{equation}
The eigenvalue equation, $f(\lambda)=0$ factorizes to the
following form:
\begin{equation}\label{eigen1}
\left(\left(G-\lambda\right)^2-\frac{1}{2}G^2A\right)
\left(\left(G-\lambda\right)^2-\frac{1}{2}G^2B\right)=0
\end{equation}
where $G=4J$, $A=1+\cos(\mathbf{q}.\mathbf{a})$ and
$B=1+\cos(\mathbf{q}.(\mathbf{c}-\mathbf{b}))$, giving
\begin{eqnarray}
\label{eigen2}
&~&\lambda(\mathbf{q})_{\pm}=4J\left(1\pm\cos\left(\frac{\mathbf{q}.\mathbf{a}}{2}\right)\right)\\
\label{eigen3}
&~&\lambda(\mathbf{q})_{\pm}=4J\left(1\pm\cos\left(\frac{\mathbf{q}.(\mathbf{c}-\mathbf{b})}{2}\right)\right)
\end{eqnarray}

One can see that there are planes in reciprocal space
perpendicular to $\mathbf{a}$ and to $(\mathbf{c}-\mathbf{b})$ for
which $\lambda(\mathbf{q})=0$. We have verified that this is not
the case for other selected $q=0$ ground states and it is these
soft modes that provide the entropic force for the final ground
state selection.

Using the mode counting arguments of Ref.~\cite{CHS} we can
determine the effect of these zero modes on the specific heat. For
a system with no soft modes, each degree of freedom will
contribute $1/2$, in units of $k_B$ to the specific heat. In the
harmonic approximation, the soft modes contribute zero. Assuming
that this singularity is regularised at higher order by  quartic
corrections, each soft mode will contribute $(1/4)$.

\begin{figure}
\centering
\includegraphics[width=0.5\textwidth]{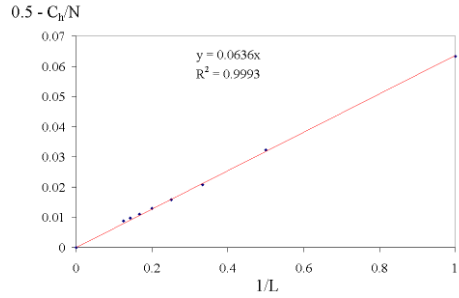}
\caption{$\frac{1}{2}-\frac{C_h}{N}$ vs $\frac{1}{L}$ for
State~1.}\label{sexy}
\end{figure}

One plane in reciprocal space contains $N_{cell}^{2/3}$ zero
modes. The number of zero modes per unit cell is
$2N_{cell}^{-1/3}$, since there are two planes. The specific heat
per spin is thus:
\begin{eqnarray}
\frac{C_h}{Nk_BT}&=&\frac{1}{4}\left[\left(4-2N_{cell}^{-1/3}\right)
\frac{1}{2}+2N_{cell}^{-1/3}\frac{1}{4}\right]\nonumber\\
&=&\frac{1}{2}-\frac{1}{8}\frac{1}{N_{cell}^{1/3}}\nonumber\\
&=&\frac{1}{2}-\frac{1}{2^{11/3}}\frac{1}{L}\nonumber\\
\frac{C_h}{Nk_BT}&=&\frac{1}{2}-C\frac{1}{L}
\end{eqnarray}
where $N_{cell}=4L^3$ and C is a constant. If
$\frac{1}{2}-\frac{C_h}{N}$ is plotted versus $1/L$, a straight
line graph should result. This figure has been tested by Monte
Carlo (see figure~\ref{sexy}). Data were collected for $L=1-7$ at
a temperature of $T/J=0.0001$ with $1,000,000$ Monte Carlo steps
per spin and $500,000$ equilibration steps, with five separate
simulations averaged for each lattice point. There is good
agreement with a linear fit, suggesting that our order by disorder
scenario is correct. However the gradient, while it is the correct
order of magnitude, is not correct: $0.063$ as opposed to the
predicted value $0.0787$. We have not, at present been able to
explain this discrepancy. Note that the entropic contribution to
the free energy coming from the soft modes scales as $N^{2/3}$ and
is therefore not extensive. This could mean that the ordering
within the $q=0$ manifold occurs at a system size dependent
temperature that goes to zero in the thermodynamic limit and the
discrepancy could come from this effect; although the temperature
here is very low. More work is required to resolve this point.

%
%
%

\section*{Conclusion}

In this paper we have reported {\it another} example of order by
disorder in a geometrically frustrated magnetic system (see for
example~\cite{CHS,spinliquid,Zit}). We have exposed one order by
disorder mechanism leading to the low temperature selection of an
ordered state, but there are several interesting statistical
mechanics questions that remain open: firstly, is there a
different, perhaps discrete mechanism, analogous to that in
Villain's original paper~\cite{Villain}, that drives the phase
transition, or are the soft modes felt by the system, even in the
disordered phase? Secondly, are there disordered states with the
same soft mode structure? This is the case for the Heisenberg
antiferromagnet on the kagom\'e lattice~\cite{CHS,Zit} and the
results here suggest that columnar line defects could exist,
equivalent to the line defects of the kagom\'e lattice. It would
be interesting to pursue this question further.

The project, in this case was strongly motivated by experimental
results on Er$_2$Ti$_2$O$_7$ and its real interest lies in the
question: is it in fact relevant? There are several details that
the simple model does not correctly predict; most notably,
experimentally the transition is second order~\cite{erbiumRC}
while here we find a very strongly first order transition.
Further, experimentally the low temperature specific heat varies
like $T^3$~\cite{erbiumRC}, which is indicative of a spin wave
spectrum with a quadratic density of energetic states. We have
calculated the density of eigenvalues above the zero frequency
modes and find $g(\lambda)=const.$, which is compatible with
having degenerate planes in reciprocal space and variations along
only one dimension. Here, of course, we cannot capture the quantum
fluctuations as our model only has one degree of freedom per spin.
To study this question more closely one would need to add out of
plane fluctuations, although it is difficult to see how this could
lead to a quadratic density of states. It might be interesting to
look at this problem in the future. However, despite these
failings our simple model does have one great success: it does
predict the correct magnetic structure for
Er$_2$Ti$_2$O$_7$~\cite{erbiumRC}. This seems particularly
interesting considering that dipolar corrections, which one might
expect to be the most substantial perturbation~\cite{BG} would
lead to a different magnetic structure~\cite{Palmer}. A classical
model, with dipoles would predict the ground state shown in
figure~\ref{GS2}, which is different from the experimentally
observed structure. With this result we suggest that our model is
a relevant starting point to study this compound and it will be
interesting to see if quantum fluctuations can change the details
discussed above.

\section*{Acknowledgments}
It is a pleasure to thank S.T. Bramwell and M. J. Harris for the collaboration from which this work stems. In
addition we have enjoyed useful discussions with B. Canals, M.J.P.
Gingras, C. Lacroix and A.S. Wills.


\section*{References}
\begin{thebibliography}{99}

\bibitem{erbiumRC}
Champion J D M \etal 2003 {\it Phys. Rev. B} {\bf 68} 020401(R)

\bibitem{Blote}
Bl\"{o}te H W J {\etal} 1969 {\it Physica} {\bf 43} 549

\bibitem{Rosenkranz} Rosenkranz S {\etal} 2000
{\it J. Appl. Phys.} {\bf 87} 5914


\bibitem{Siddharthan}
Siddharthan R {\etal} 1999 {\it Phys. Rev. Lett.} {\bf 83} 1854

\bibitem{MMM}
Harris M J {\etal} 1998 {\it J. Magn. Magn. Mater.} {\bf 177} 757

\bibitem{Villain}
Villain J 1980 {\it J. Physique} {\bf 41} 1263

\bibitem{RBS}
Reimers J N, Berlinsky A J and Shi A C 1991 {\it Phys. Rev. B}
{\bf 43} 865

\bibitem{BOAS} Boas M L 1983 {\it Mathematical Methods in the Physical
Sciences} (New York, Chichester: Wiley)


\bibitem{PHD}  Champion J D M 2001 Ph.D. Thesis, University of London

\bibitem{BGR}
Bramwell S T, Gingras M J P and Reimers J N 1994 {\it J. Appl.
Phys.} {\bf 75} 5523

\bibitem{CHS}
Chalker J T, Holdsworth P C W and Shender E F 1992 {\it Phys. Rev.
Lett.} {\bf 68} 855






\bibitem{spinliquid}
Moessner R and Chalker J T 1998 {\it Phys. Rev. Lett.} {\bf 80}
2929

\bibitem{Zit}
Zhitomirsky M E 2002 {\it Phys. Rev. Lett.} {\bf 87} 057204

\bibitem{BG} Bramwell S T and Gingras M J P 2001 {\it Science}
{\bf 294} 1495

\bibitem{Palmer}
Palmer S E and Chalker J T 2000 {\it Phys. Rev. B} {\bf 62}, 488


\end{thebibliography}
\end{document}